\documentclass[12pt,preprint]{aastex}
\renewcommand{\cite}{\citealp}

\shorttitle{Variable stars in B514}
\shortauthors{Clementini et al.}

\begin{document}

\title{The Variable Star Population of the globular cluster B514 in the Andromeda 
Galaxy\altaffilmark{*}}
\author{ Gisella Clementini\altaffilmark{1}, Rodrigo Contreras\altaffilmark{1,2},
Luciana Federici\altaffilmark{1}, Carla Cacciari\altaffilmark{1}, Roberto Merighi\altaffilmark{1}, Horace A. 
Smith\altaffilmark{3}, M\'arcio Catelan\altaffilmark{4,5,}, Flavio Fusi Pecci\altaffilmark{1}, Marcella 
Marconi\altaffilmark{6}, Karen Kinemuchi\altaffilmark{7,8}, and Barton J. 
Pritzl\altaffilmark{9}}
\altaffiltext{1}{INAF, Osservatorio Astronomico di Bologna, Bologna, 
Italy; gisella.clementini@oabo.inaf.it 
}
\altaffiltext{2}{Dipartimento di Astronomia, Universit\`a di Bologna, Bologna, Italy}
\altaffiltext{3}{Department of Physics and Astronomy, Michigan State
University, East 
Lansing, MI 48824, USA; kuehncha@msu.edu, smith@pa.msu.edu, beers@pa.msu.edu}
\altaffiltext{4}{Pontificia Universidad Cat$\rm{\acute{o}}$lica de Chile, Departamento de Astronom\'{\i}a y Astrof\'{\i}sica, 
Santiago, Chile; mcatelan@astro.puc.cl}
\altaffiltext{5}{John Simon Guggenheim Memorial Foundation Fellow} 
\altaffiltext{6}{INAF, Osservatorio Astronomico di Capodimonte, Napoli, Italy; marcella@na.astro.it}
\altaffiltext{7}{Universidad de Concepci\'on, Departamento de Astronom\'{\i}a, Concepci\'on, Chile, kkinemuchi@astro-udec.cl}
\altaffiltext{8}{Department of Astronomy, University of Florida, 211 Bryant Space Science Center, Gainesville, FL 32611-2055, USA; 
kinemuchi@astro.ufl.edu}
\altaffiltext{9}{Department of Physics and Astronomy, 
University of Wisconsin Oshkosh, Oshkosh, WI 54901, USA; pritzlb@uwosh.edu}
\altaffiltext{*}{Based on data collected with the Wide Field Planetary Camera 2 on board of the
Hubble Space Telescope and ACS HST archive data.}

\begin{abstract}
A rich harvest of RR Lyrae stars has been identified for the first time in B514, a 
metal-poor ([Fe/H]$\sim$ --1.95 $\pm$0.10 dex) globular cluster of the Andromeda galaxy (M31),  
based on Hubble Space 
Telescope Wide Field Planetary Camera 2 and Advanced Camera for Surveys time-series 
observations. We have detected and derived periods for 89 RR Lyrae stars (82 fundamental-mode $-$RRab$-$ and 
7 first-overtone $-$RRc$-$ pulsators, respectively) among 161 candidate variables identified in the cluster. 
The average period of the RR Lyrae variables ($\langle Pab \rangle$ = 0.58 days 
and $\langle Pc \rangle$ = 0.35 days, for RRab and RRc pulsators, respectively) and the position in the 
period-amplitude diagram both suggest that B514 is likely an Oosterhoff type I 
cluster. This appears to be in disagreement with the general behaviour of the metal-poor globular 
clusters in the Milky Way, which show instead Oosterhoff type II pulsation properties.

The average apparent magnitude of the RR Lyrae stars sets the mean level of the cluster 
horizontal branch at $\langle V(RR) \rangle = 25.18 \pm 0.02$ 
($\sigma$=0.16 mag, on 81 stars). By adopting a reddening $E(B-V) = 0.07 \pm 0.02$ mag 
(\citealt{sc98}), the above metallicity and M$_V$=0.44$\pm$0.05 mag for the RR Lyrae variables 
of this metallicity (\citealt{clem03}), 
we derive a distance modulus of $\mu_{0}=24.52 \pm 0.08$ mag, corresponding 
to a distance of about $800 \pm 30$ kpc, based on a value of M$_V$ that sets 
 $\mu_{0}$(LMC)=18.52.

\end{abstract}


\section{Introduction}
The Andromeda galaxy provides us with a unique opportunity to study the 
formation and  evolution of a massive spiral galaxy other than  the Milky Way
(MW). 
Various authors (\citealt{vdb00,vdb06}) suggested that M31 originated as an 
early merger of two or more massive metal-rich progenitors, accounting for the galactic halo 
wide range in metallicity (\citealt{dhp01}) and age (\citealt{bro03}) compared to the MW.
There is also an interesting suggestion by \citet{kra02} that the ``young'' second-parameter 
globular clusters in our Galaxy, as well as at least some of the MW dwarf spheroidal companions, 
were in fact accreted 
from M31 when Andromeda was forming its Population II stars.

The pulsation properties of the variable stars in the field and globular
clusters (GCs) of M31 have the potential
to provide essential insights on the galaxy formation and to trace the
merging episodes that led to the early assembling of the galaxy.
The RR Lyrae stars, in particular, belonging to the old stellar population 
(t $> 10$ Gyr), were eyewitnesses to the first epochs of star formation in M31. 
Their mean periods along with the metallicities of the parent clusters can be used to 
set the luminosity of the horizontal branch (HB), and hence can provide 
precious information on how the M31 halo formed and evolved, in comparison with their 
MW counterparts (see e.g. \citealt{cat04}, 2009).

In the MW the vast majority of GCs which contain significant numbers of 
RR Lyrae stars sharply divide into two very distinct classes, the Oosterhoff
types I and II (\citealt{oo39}; see \citealt{cat09} for a recent review) according 
to the mean 
pulsation periods of their RR Lyrae variables. Oosterhoff type I (OoI) GCs have 
$\langle Pab \rangle \simeq 0.56$ days, 
Oosterhoff type II clusters (OoII) have $\langle Pab \rangle \simeq$ 0.66 days (\citealt{cle01}). 
This phenomenon is referred to as the Oosterhoff dichotomy. 
There is evidence that OoI and OoII clusters in the MW may have different 
kinematical and spatial distributions thus possibly resulting from different 
accretion or formation events in the halo. This is supported also by a 
difference in mean chemical abundance, OoII clusters being on average more 
metal-poor than OoI clusters, and possibly by a difference in age, metal-poor
clusters being on average slightly older than the intermediate-metallicity
ones (\citealt{vdb93};  \citealt{dea05}; \citealt{mar09}).
Whatever the mechanism, it is clear 
that the Oosterhoff dichotomy reflects conditions within the MW halo at the 
time of GC formation.
The existence or absence of the Oosterhoff phenomenon among the  M31 GCs  
therefore provides  information on the halo formation processes  and thus on 
the chemical/dynamical  evolution history of the dominant galaxy of the 
Local Group (\citealt{cat04}, 2009). 
Detection and characterization of RR Lyrae stars in M31 globular clusters 
is challenging since they are too crowded to be well-studied 
with ground-based telescopes. 
There has been only one previous attempt
at detecting RR Lyrae stars in the globular clusters of M31.
Clementini et al. (2001) used archival data obtained with the Wide Field Planetary Camera 2 
(WFPC2) onboard the Hubble Space Telescope 
(HST) to make the first tentative detection of RR Lyrae stars in 
4 clusters, namely, G11, G33, G64, and G322.
A number of RR Lyrae candidates were identified in each cluster (two, four, eleven,  and eight variables, 
in G11, G33, G64, and G322, respectively), but 
the small number of available data points and the short time baseline did not 
allow a definition of light curves and hence periods.
As part of a Cycle 15 HST program we have observed a number of M31 GCs to properly characterize their variable star population.
The clusters were selected as to have metallicities that, in the Milky Way, would place them either in the OoI or in the OoII groups.
Here we present results 
for B514, a globular cluster 
located at a projected distance 
of $\sim55$ Kpc from the M31 center, not far from the galaxy's major axis 
(\citealt{gal05}). The color-magnitude diagram (CMD) indicates that B514 
is a classical, old metal-poor globular cluster.  The metallicity [Fe/H] is estimated as the 
weighted mean value of a few independent determinations, namely --1.8$\pm$0.3 
(spectroscopic from \citealt{gal05}), 
--1.8$\pm$0.15 (from the CMD, \citealt{gal06}), and --2.14$\pm$0.15 (from the CMD, \citealt{mck07}). 
The integrated absolute magnitude 
$M_V$ has been estimated of $\sim -9.1$ mag, and classifies the cluster among the 
brightest globulars of M31. Moreover, the observed half-light radius $r$ = 1.6 arcsec, corresponding to 
$\sim$ 6 pc at the distance of M31 (see below), is significantly larger than for most clusters 
of the same luminosity (\citealt{fed07}). 
%

In this Letter we present results of a search for variable stars 
which lead us to the discovery of about a hundred RR Lyrae stars in B514.
This number is larger than found in the vast majority of the MW globular clusters, and since 
we are not able to resolving the cluster's core, there are likely many more RR Lyrae stars
in B514.
This is the first time that a large sample of RR Lyrae stars is discovered in an M31 globular cluster
and that their pulsation properties (periods, amplitudes, etc.) are fully characterized
thus allowing to establish the cluster's Oosterhoff type. 
Here we focus on the Oosterhoff classification of B514 and on its importance to get hints on the 
formation of the Andromeda galaxy by comparison with the properties of variables in the MW globular clusters.
%
%
We also present a $V, V-I$ CMD of B514 based on the ACS data extending 
to $V = 28$ mag. This is about half a magnitude fainter than the cluster CMDs  by \citet{gal06} and 
\citet{mck07}. 
The complete list of the cluster variables with ephemerides, photometric data and light curves, will be presented 
in Contreras et al. (2009, in preparation), where we will discuss in  more detail their specific 
pulsation and evolutionary properties.

\section{Observations and Data Reduction}
Time-series F606W, F814W observations of B514 (R.A.= 00$^h$ 31$^m$ 09$^s$.83, 
decl.= 37$^{\circ}$ 53$^{\prime}$ 59$^{\prime \prime}$.6) were obtained with 
the WFPC2 under HST program GO
11081 (PI G. Clementini) in June 12-14, 2007.
The time series consisted of fifteen individual exposures in each filter, of 1100$^s$ length, 
taken by alternating the filters. 
They were scheduled in three 10-orbit blocks executed almost continuously from 2007,  
June 12, UT 18$^h$ 16$^m$ 19$^s$, to  2007, June 14, UT 01$^h$ 23$^m$ 38$^s$, for total exposure times of 4 hours and 35 min in each band.  
The WFPC2 proprietary data were combined with ACS/WFC archive data from program GO 10394 (PI  N.R. Tanvir),
covering the period 2005, July 19-20, and from program GO 10565 (PI S. Galleti)
obtained in 2006, June 10, making a total of 20 phase-points 
in the F606W-band and 21 phase-points in the F814W-band.
Photometric reduction of the individual pre-reduced images supplied by the STScI
pipeline was performed using HSTPHOT and Dolphot/ACSPHOT (\citealt{dol00a,dol00b}). These are PSF-fitting photometry packages 
specifically designed to perform photometry on point sources 
observed with HST/WFPC2 and HST/ACS. The packages identify sources above a fixed flux threshold and perform
photometry on individual frames, taking into account information on CCD defects that are 
attached to the pre-reduced frames and automatically applying the correction for charge transfer efficiency (CTE) (\citealt{dol00a}). 
Position, flight system magnitudes and Johnson-Cousins standard magnitudes of the sources are provided as output, 
along with a number of quality parameters of the measured sources (see \citealt{dol00a}, for details). 
%
%
%
Typical errors of the combined photometry (8 ACS frames) for non-variable stars
at the magnitude level of the B514 horizontal branch (HB, $V \sim 25.2$ mag)  are
$\sigma_{V}$ = 0.017 mag, $\sigma_{I}$= 0.015 mag for the ACS dataset, 
and $\sigma_{V}$ = 0.019 mag,  $\sigma_{I}$= 0.032 mag for the WFPC2 data. 

\section{Variable Stars}
Variable star candidates were identified from the scatter diagrams of the F606W and F814W datasets, using  VARFIND, 
custom software developed at the Bologna Observatory by 
P. Montegriffo.  
We first analyzed the WFPC2 proprietary data, whose extended time series are optimized for the detection of
variables of RR Lyrae type. The WFPC2 candidate variables were then counteridentified on the ACS data.
A further search with VARFIND  was later performed on the ACS archive data, and the counteridentification with the WFPC2 data was iteratively repeated.
The search procedure returned a final catalogue of 161 candidate variables, most of which are located 
on the cluster HB. Many more will likely exist in the core of the cluster, that we have not detected. 
Periods and classification in type of the candidate variables were derived from the study of the 
light curves with Graphical Analyzer of Time Series (GRaTiS; see \citealt{clem00}).
%
%
We obtained reliable periods and light curves for 89 RR Lyrae stars: 82 fundamental-mode 
(RRab) and 7 first-overtone (RRc) pulsators.  
For the remaining 72 candidates we still lack a firm classification, mostly because of scatter 
in the data.  
The time sampling of our data covers nearly 29 hours in three observing blocks of 7 hours 
each, separated by two gaps of 50 minutes and 7 hours respectively. In addition, we have 
6 epochs of archive data spaced by 694 and 368 days. This ensures that all possible periods 
of RR Lyrae stars are well sampled and no bias or alias effects are present, and allows 
us to derive accurate periods mostly to the 4th decimal digit.

\begin{figure*}
\includegraphics[width=16.3cm,clip]{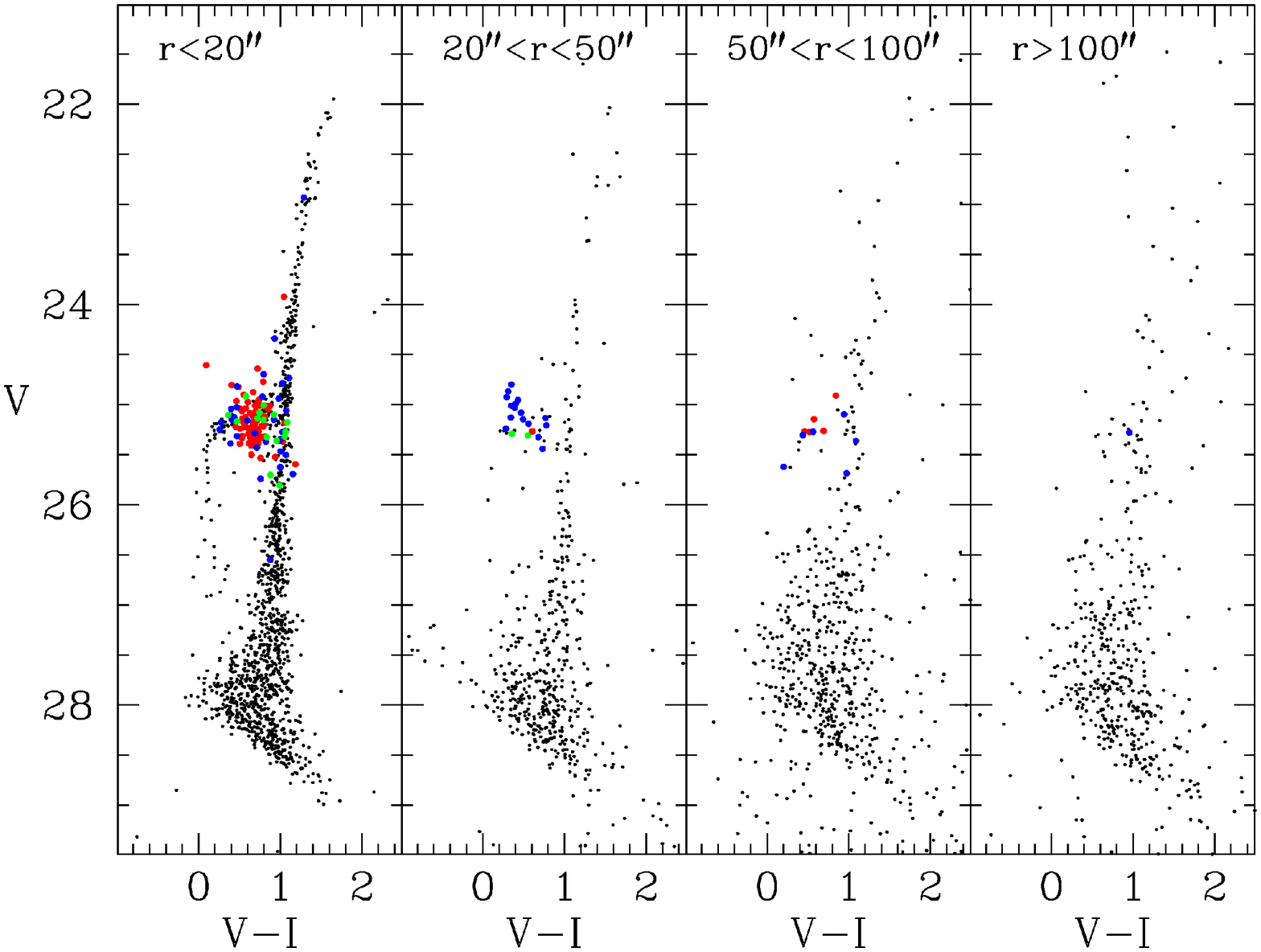}
\caption{$V, V-I$ CMD of B514 in 4 regions with increasing the distance from the cluster 
center which was set at R.A.= 00$^h$ 31$^m$ 09$^s$.83, 
decl.= 37$^{\circ}$ 53$^{\prime}$ 59$^{\prime \prime}$.6  (J2000), based on our reductions of the ACS 
archive photometry.  
Bona-fide RR Lyrae
stars are marked by filled red circles, candidate variable stars by blue and green filled circles (see Section 3).
}
\label{f:figcmd}
\end{figure*}

Figure~\ref{f:figcmd} shows the $V,V-I$ CMD of B514, 
obtained 
from the ACS data, for stars in four separate regions at increasing distance from the cluster center.
Only sources with
object type flag=1 (i.e. best measured stars), crowding $<$0.3, 0.5$<$sharpness$<$0.5,
$\chi^2<$1.5 for V$>$ 24 mag and $\chi^2<$2.5 for brighter stars are shown in the 
figure\footnote{The 
photometric data generating the CMD are available in electronic form upon request.}. 
Confirmed variable stars are plotted as red filled circles, candidate variables 
with sufficient light curve sampling as blue filled circles, and less reliable candidates as green filled circles.

\begin{figure*}
\centering
\includegraphics[width=13.5cm,clip]{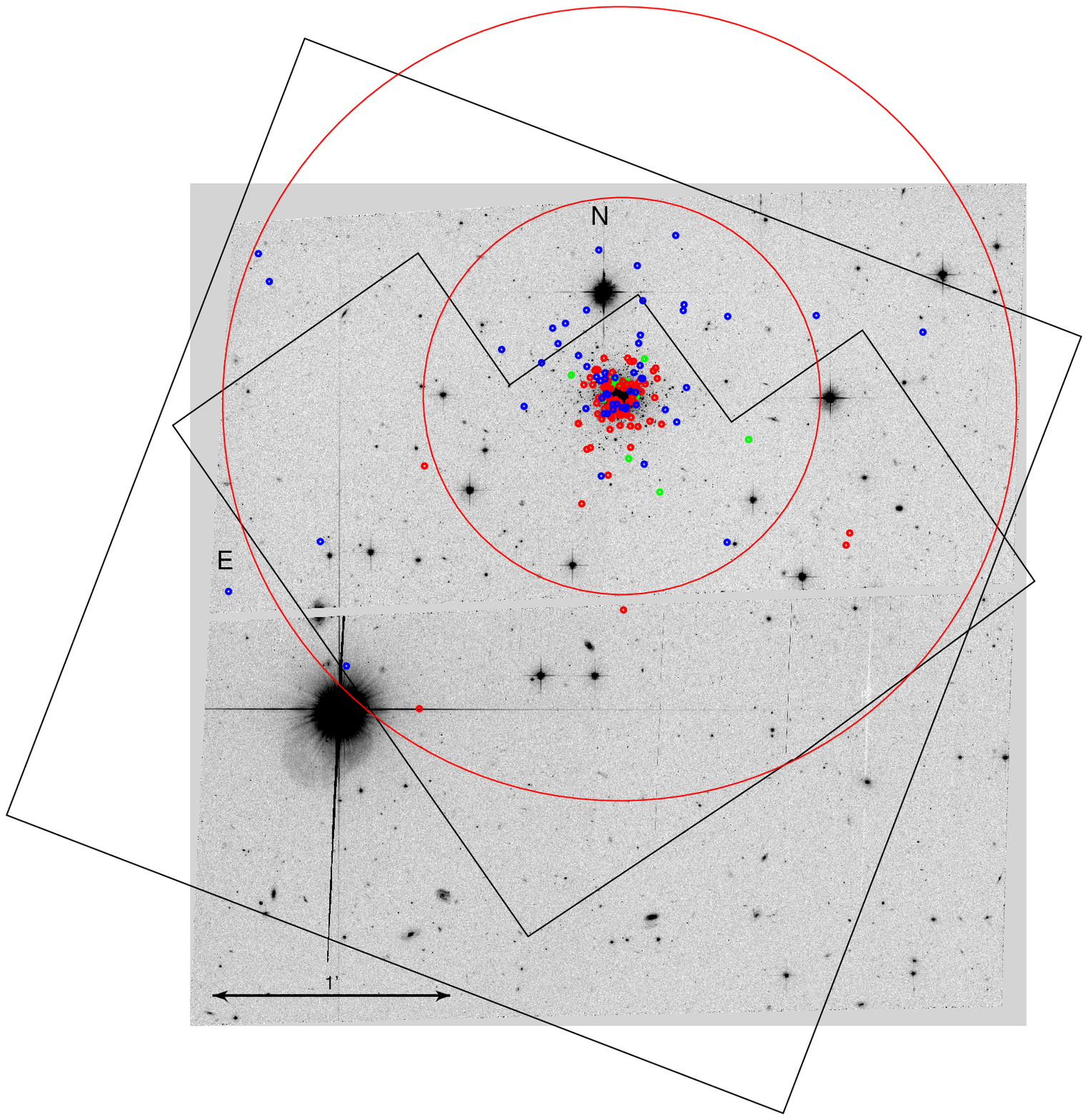} 
\includegraphics[width=8.5cm,clip]{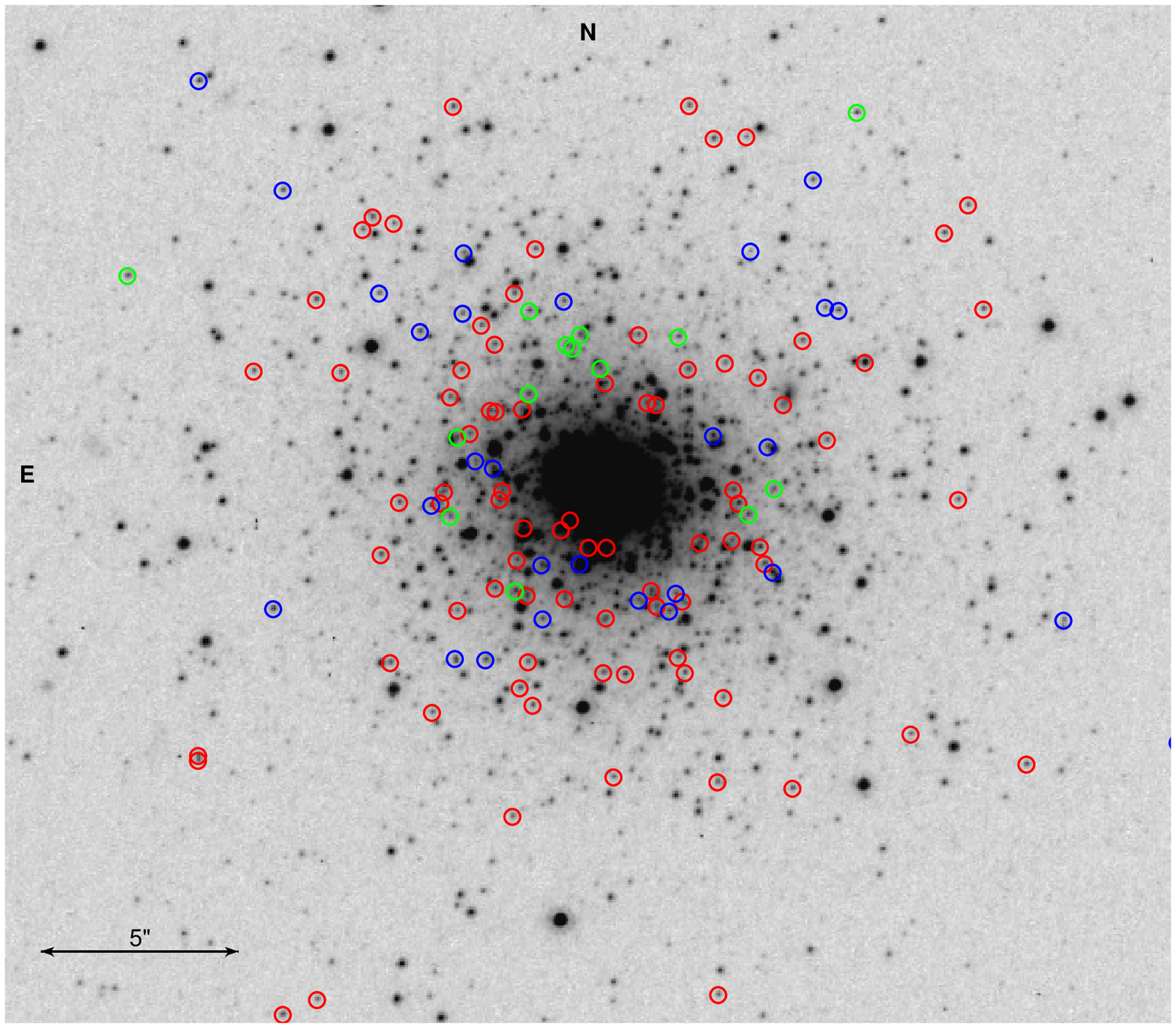}
\caption{{\it Upper panel:} Location of B514 confirmed (red circles) and candidate (blue and green circles) variable stars 
on a  $3^{\prime}$ 22$^{\prime \prime} \times 3^{\prime}$ 22$^{\prime \prime} $ map of the cluster, corresponding to the field of 
view covered by the ACS observations of HST GO 10565 (see \citealt{gal06}). Also shown are the field of 
view covered by the ACS observations of HST GO 10394 (see \citealt{mck07}; tilted square), and the field of view of our 
WFPC2 observations. The two red circles correspond to radii of $r = 50^{\prime \prime}$ and $r = 100^{\prime \prime}$ from 
the center of B514.
{\it Lower panel:}  Enlargement of the variable stars' map showing a 
$30^{\prime \prime} \times 30^{\prime \prime}$ region centered on the cluster. 
}
\label{f:figisto}
\end{figure*}

\begin{figure*} 
\includegraphics[width=16.3cm,clip]{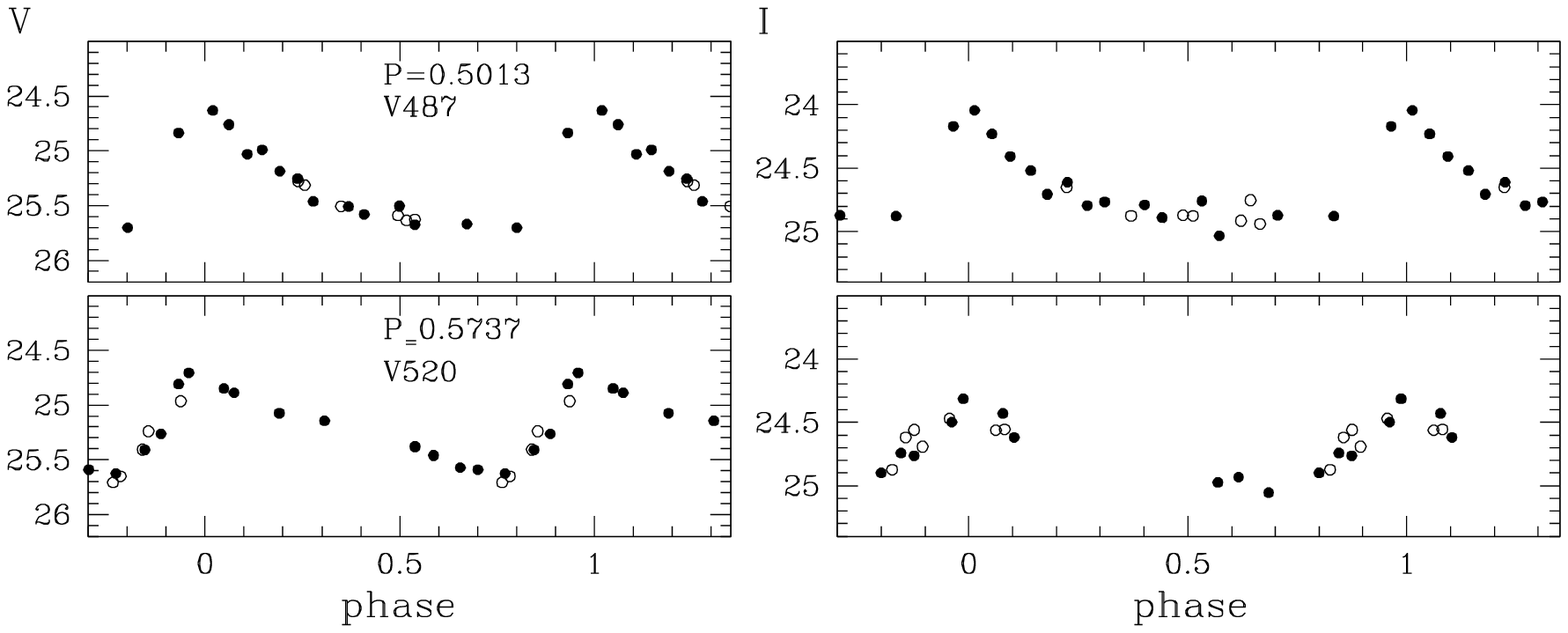}
\includegraphics[width=16.3cm,clip]{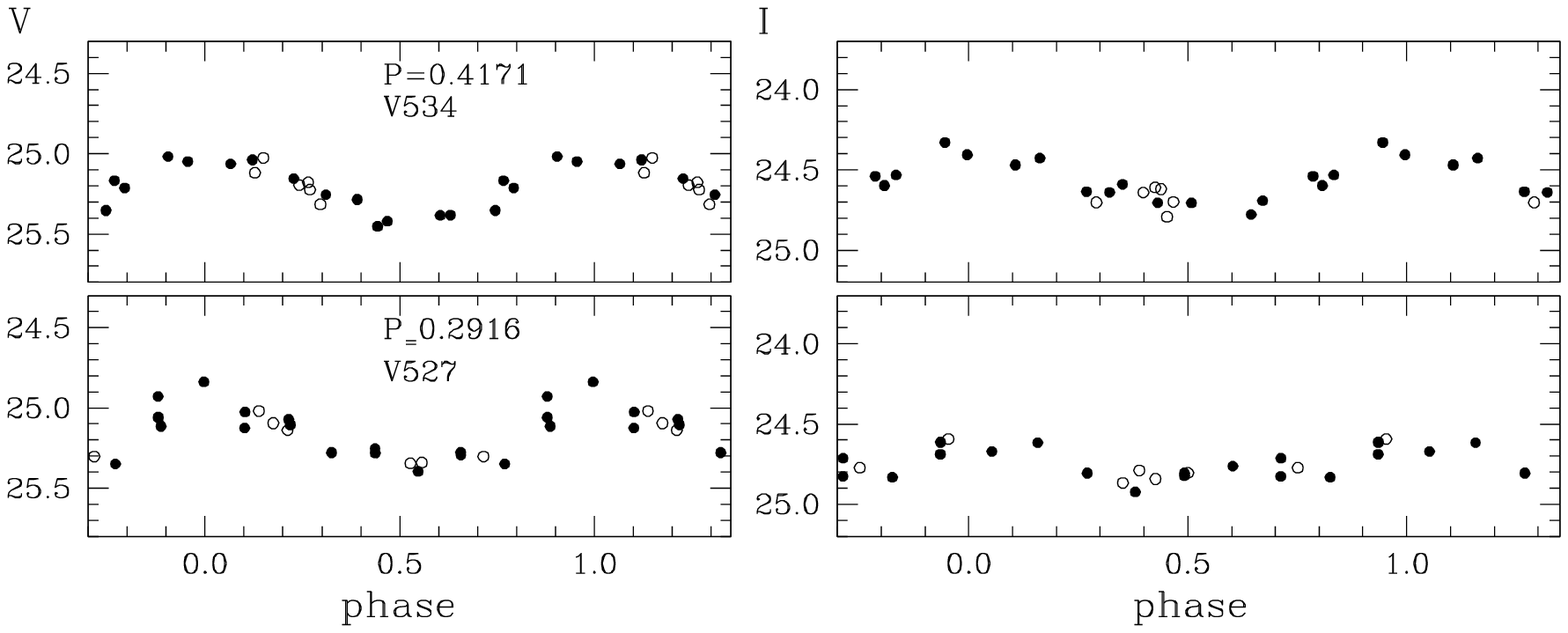}
\caption{$V$ (left panels) and $I$ (right panels) light curves of RR Lyrae stars identified in B514.
{\it Two upper rows:} fundamental-mode pulsators: {\it Two lower rows:} first-overtone pulsators. 
Filled and open circles indicate WFPC2 and ACS data, respectively. Typical error of the single data point 
at the magnitude level of the HB is about 0.06 mag. }
\label{f:figlc}
\end{figure*}

According to their 
position on the CMD, the vast majority of the candidate variables discovered in B514 are likely RR Lyrae stars.
However, our candidate's list also includes a number of binaries and a few objects above and below the HB.
Variable stars are plotted in the CMD using mean magnitudes and colors computed as simple averages of the available photometric measurements.
The scatter observed around the average level of the HB is therefore largely due to variable stars 
with uneven coverage of the light curve, and  will likely be significantly reduced when the mean 
magnitudes will be derived averaging over the full pulsation cycle. 
The upper panel of Figure~2 shows the location of the B514 variable 
stars on the $3^{\prime}$ 22$^{\prime \prime}\times 3^{\prime}$ 22$^{\prime \prime}$ 
field of view covered by the ACS observations of \citet{gal06}. The lower panel of Figure~2 shows an 
enlargement of the variable stars' map corresponding to  a $30^{\prime \prime} \times 30^{\prime \prime}$ region around the cluster center.
As noted in \citet{fed07}, B514 is rather extended (see their Fig. 6) and its tidal radius was estimated as $\sim$ 17 arcsec 
($\sim$ 65 pc) from the analysis 
of the distribution of the integrated light and star counts. 
However, we have detected variables well beyond this distance, in particular five RR Lyrae stars are 
located farther than 50 arcsec and as far as $\sim$ 90 arcsec from the cluster 
centre (Fig. \ref{f:figisto}). Although they are likely field stars, 
the CMD shown in Fig. \ref{f:figcmd} indicates that a non negligible fraction of the cluster 
population is still present in the annulus 50-100 arcsec, suggesting the possibility that at 
least a few of these distant variables may belong to the cluster. 
Radial velocity membership will be the most reliable way to assess to which population they belong.   
Examples of  light curves for 2 fundamental-mode and 2 first-overtone RR Lyrae stars we have identified 
in B514 are shown in Figure~\ref{f:figlc}. 
The mean periods of the 89 confirmed RR Lyrae stars in 
B514 are $\langle Pab \rangle \sim$0.58 days and $\langle Pc \rangle \sim$ 0.35 days
for fundamental mode and first-overtone variables, respectively, suggesting a 
possible dominance of Oosterhoff type I.
Figure \ref{f:figpa} shows the $V$-band period-amplitude diagram of the B514 RR Lyrae 
stars along with the Oosterhoff loci defined by the Galactic globular clusters from 
\citet{cle00} (linear relations), and the period-amplitude distributions of 
\emph{bona fide} regular and  evolved RRab stars in M3 (quadratic relations), 
from Cacciari et al. (2005). 

The B514 RR Lyrae stars are  close to the loci of 
Oosterhoff type I systems and regular RRab stars in M3.
In conclusion, B514 appears to be a somewhat borderline OoI cluster, as it seems to follow 
a different rule than what is found in the MW where metal-poor ([Fe/H]$\lesssim-1.7$) GCs containing 
RR Lyrae stars show Oosterhoff type~II properties. 
 However, these results should be taken with caution since B514 could be a peculiar cluster. The entire sample 
of M31 clusters observed in our Cycle 15 program needs to be 
carefully analyzed (see Contreras et al. 2010, in preparation) to reach more firm conclusions 
on the Oosterhoff properties of the M31 globular clusters.  


\section{CMD and Distance}

The CMD of B514 (see Figure~1) reaches $V \sim $ 28 mag and is very rich in stars. 
It has well-populated  
horizontal and red giant branches.  The HB stretches across the RR Lyrae 
instability strip, which is entirely
filled by the large number of confirmed and candidate RR Lyrae stars, and extends 
significantly to the blue through a blue tail that reaches $V \sim 27$ mag at $B-V \sim $0.1 mag. 
The red giant branch (RGB) is a prominent feature of the 
CMD, and does not exhibit significant scatter, thus ruling out a metallicity spread in B514.
The mean magnitude of the B514 RR Lyrae stars is 
${\rm \langle V\rangle}$ = 25.18 $\pm 0.02$ mag 
(with a dispersion of 0.16 mag among the 81 stars). We adopt an absolute 
magnitude of $\rm{M_{V}}=0.44\pm0.05$ mag for RR Lyrae stars with the metallicity 
$\rm{[Fe/H]}=-1.95\pm0.1$ dex of B514 \citep{clem03}, which is consistent with an  
LMC distance modulus of 18.52. Using the reddening  value 
of $E(\bv)$=0.07$\pm$0.02 estimated from \citet{sc98},  
we find a distance modulus of $\mu_{0}=24.52\pm 0.08$ mag, 
which corresponds to a distance of about 800 $\pm$ 30 kpc, in good agreement  
with the distance 
adopted for this cluster by \citet{gal06}.
%
%

\section{Conclusions}
\begin{figure}
\includegraphics[width=16.3cm,clip]{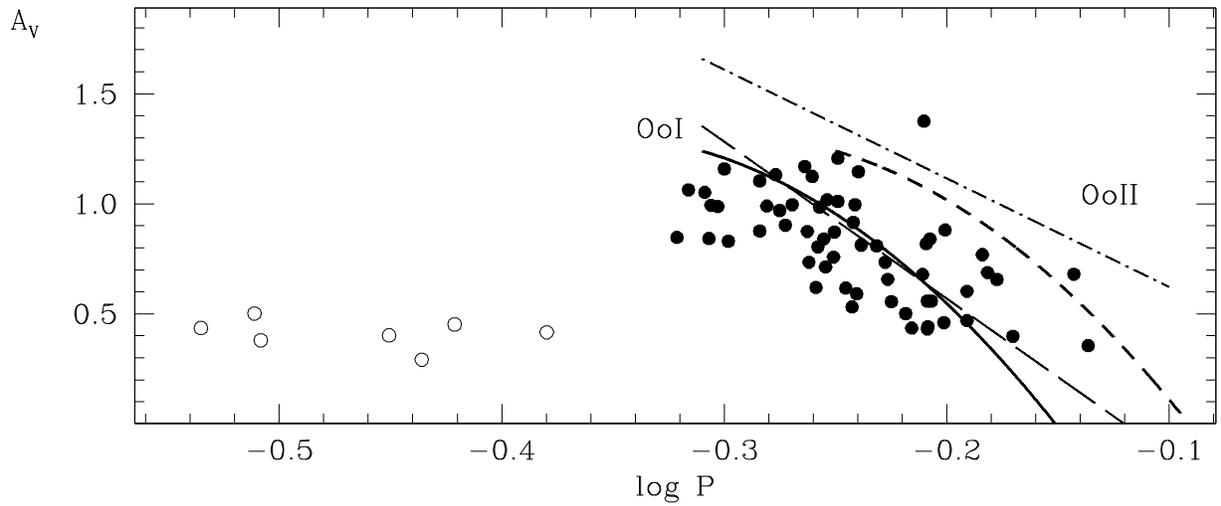}
\caption{Period-amplitude diagram in the $V$ band. RRab variables are shown by 
filled circles, RRc stars by open circles. 
The linear relations show the period-amplitude distributions of 
Galactic Oosterhoff type I and II clusters (\citealt{cle00}).   
The quadratic relations show the \emph{bona fide} regular 
and  evolved RRab stars in M3, from \citet{cac05}, for comparison; 
analogous relations for the RRc stars are not reported 
because of the small number of these stars so far detected in B514.  
}
\label{f:figpa}
\end{figure}

We have identified and obtained periods and light curves for 82 RRab stars and 7 RRc 
stars in the globular cluster B514 of the Andromeda galaxy. 
The average period of the B514 RRab stars for which we have complete and reliable 
light curves and their location on the period-amplitude diagram 
indicate likely OoI-type characteristics (see right side of Fig. \ref{f:figpa}), 
however the cluster's  low metallicity is more typical of an OoII-type. 
%
Thus B514 seems to follow a different rule than what is found in the MW, where 
metal-poor ([Fe/H]$\lesssim-1.7$) GCs containing RR Lyrae stars have Oosterhoff II type.
This may suggest that B514 is indeed a somewhat peculiar cluster, as indicated by 
independent evidence of some similarity with peculiar clusters in the MW such 
as $\omega$ Cen, M54 and NGC2419 (\citealt{fed07}). 
Alternatively, we are seeing an indication that the M31 GCs have different RR Lyrae pulsation 
characteristics than that seen in the main body of the MW GCs. 
 A more detailed analysis will be done  based on the entire sample 
of M31 clusters observed in our Cycle 15 program, and the results will be presented  in a 
forthcoming  paper (Contreras et al. 2010).
%

\bigskip
\acknowledgments  
We warmly thank the Program Coordinator A. Roman, and the Contact Scientist M. Sirianni, of our HST program for their 
invaluable help with the Phase II and scheduling of the HST observations. This research was partially supported by 
COFIS ASI-INAF I/016/07/0.
HAS and BP thank the Space Telescope Science Institute for support under grant
HST-GO-11081.05-A.
M.C. is supported by Proyecto Fondecyt Regular 1071002, by Proyecto Basal PFB-06/2007, 
by FONDAP Centro de Astrof\'{i}sica 15010003, and by a John Simon Guggenheim Memorial Foundation 
Fellowship.



%
%

\end{document}